\documentclass[%
reprint,
superscriptaddress,
 amsmath,amssymb,
 aps,
]{revtex4-1}

\def\Eb{\mathbf{E}}
\def\ub{\mathbf{u}}
\def\Jb{\mathbf{J}}
\usepackage{xcolor}
\usepackage{graphicx}
\usepackage{dcolumn}
\usepackage{bm}
\usepackage[normalem]{ulem}
\usepackage{soul}
\usepackage{cancel}

\begin{document}

\preprint{APS/123-QED}

\title{Electric fields in liquid water irradiated with protons at ultra-high dose rates}

\author{F.Gobet}
 \email{gobet@lp2ib.in2p3.fr}

\affiliation{Univ. Bordeaux, CNRS, LP2I, UMR 5797, F-33170 Gradignan, France}%

\author{P.Barberet}
\affiliation{Univ. Bordeaux, CNRS, LP2I, UMR 5797, F-33170 Gradignan, France}%

\author{M.-H.Delville}%
\affiliation{Univ. Bordeaux, CNRS, ICMCB, UMR 5026, F-33608, Pessac, France}%

\author{G.Devès}

\affiliation{Univ. Bordeaux, CNRS, LP2I, UMR 5797, F-33170 Gradignan, France}%

\author{T.Guérin}%
\affiliation{Univ. Bordeaux, CNRS, LOMA, UMR 5798, F-33400 Talence, France}%


\author{R.Liénard}%
\affiliation{Univ. Bordeaux, CNRS, LP2I, UMR 5797, F-33170 Gradignan, France}%


\author{H.N.Tran}%
\affiliation{Univ. Bordeaux, CNRS, LP2I, UMR 5797, F-33170 Gradignan, France}%

\author{C.Vecco-Garda}%
\affiliation{Univ. Bordeaux, CNRS, ICMCB, UMR 5026, F-33608, Pessac, France}%

\author{A.W\"urger}%
\affiliation{Univ. Bordeaux, CNRS, LOMA, UMR 5798, F-33400 Talence, France}%

\author{S.Zein}%
\affiliation{Univ. Bordeaux, CNRS, LP2I, UMR 5797, F-33170 Gradignan, France}%

\author{H.Seznec}%
\affiliation{Univ. Bordeaux, CNRS, LP2I, UMR 5797, F-33170 Gradignan, France}%

\date{\today}

\begin{abstract}

We study the effects of irradiating water with 3 MeV protons at high doses by observing the motion of charged polystyrene beads outside the proton beam. By single-particle tracking we measure a radial velocity of the order of microns per second. Combining electrokinetic theory with simulations of the beam-generated reaction products and their outward diffusion, we find that the bead motion is due to electrophoresis in the electric field induced by the mobility  contrast of cations and anions. This work sheds light on the perturbation of biological systems by high-dose radiations and paves the way for the manipulation of colloid or macromolecular dispersions by radiation-induced diffusiophoresis.
\end{abstract}

\keywords{a}
\maketitle

{\it Introduction.--} Recent advances in micro-beam irradiation \cite{ialyshev2022enhancing,eling2019ultra}  as well as in electron or X-ray microscopy \cite{wu2020liquid,hemonnot2017imaging,mccoll2012caenorhabditis,smith2020liquid,sung2022liquid,chee2019direct,smith2020liquid,mirsaidov2020liquid,evans2011controlled,grogan2014bubble,wang2018longer,moser2018role}  of liquid or biological samples have revived studies of the physical, chemical and biological effects resulting from the interaction between ionizing radiation and matter at ultra-high dose rates (10 kGy s$^{-1}$ - 100 MGy s$^{-1}$). In the context of high resolution imaging with this extreme regime of irradiation, the question of whether the obtained images are representative of the original system or the disturbed one remains open \cite{mirsaidov2020liquid,sung2022liquid,chee2019direct,smith2020liquid,evans2011controlled,grogan2014bubble,wang2018longer,moser2018role}. Indeed, at ultra-high dose rates, various physico-chemical processes have recently been demonstrated in the imaged sample, such as the synthesis of nanoparticles \cite{evans2011controlled}, the generation of H$_2$ nanobubbles \cite{grogan2014bubble}, the degradation of polymers in solution \cite{wang2018longer}, or the shrinking of bacterial cells \cite{moser2018role}. Understanding and quantifying the processes involved in an irradiated liquid target is therefore crucial for any applications in high-resolution imaging of soft matter or biological systems.  In particular, the perturbation of the sample outside the applied micro-beam, through the generation of induced fields, remains an open question.

And yet, it is well known  that ionizing beams induce the radiolysis of water, and therefore the local creation of ionic species  \cite{swallow1973radiation}. One can therefore expect a gradient of ionic species around the beam, possibly generating electrostatic fields. In this Letter, we show that this effect is strong enough to induce the migration of charged colloidal beads away from the beam, enabling us to demonstrate the existence of an electric field, even at distances ten times larger than the beam diameter (or more). This observation is in semi-quantitative agreement with the predictions of a reaction-diffusion model of water radiolysis in which electrical effects are taken into account. This work creates an unexpected bridge between the fields of interactions of ionizing beams with matter and diffusiophoresis~\cite{derjaguin1947kinetic,anderson1982motion,anderson1989colloid,anderson1984diffusiophoresis},  which has been intensively studied over the last decade, following major advances in microfluidics in the context of manipulation of colloids ~\cite{shim2022diffusiophoresis,ajdari2006giant,abecassis2008boosting,paustian2015direct,shin2016size,palacci2012osmotic,sear2019diffusiophoresis,shin2017membraneless,prieve1984motion,chiang2014multi,banerjee2016soluto}. In particular, as the micro-beam can be alternatively applied to different regions of the sample, our set-up opens the possibility of producing inhomogeneous and transient electric fields to manipulate and/or trap colloids or macromolecules.

\begin{figure}[h]

\includegraphics[width=8.6cm,trim=100 200 500 250, clip=true]{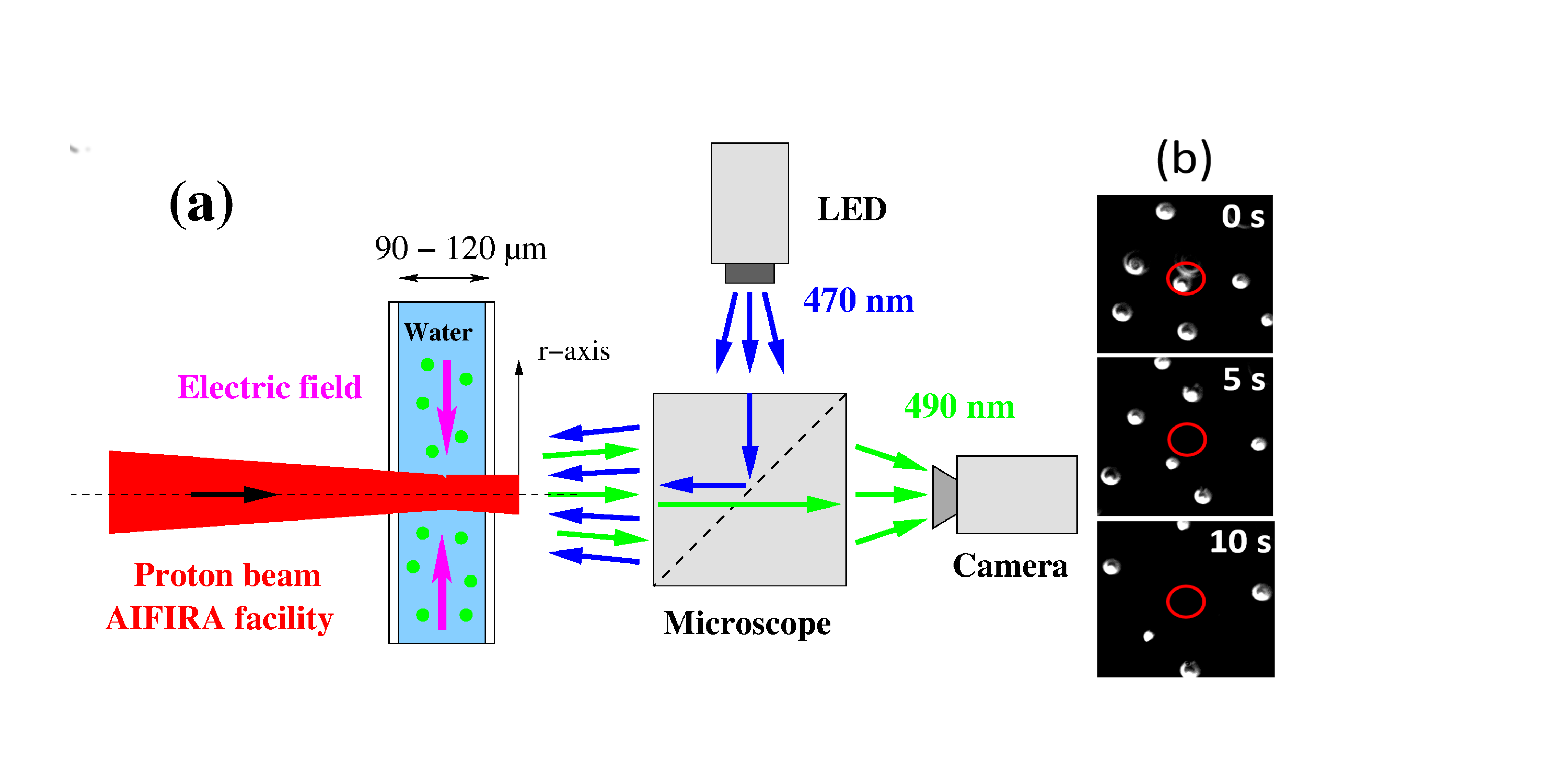}

\caption{\label{Fig1}(a) A schematic illustration of the experimental setup. The pink arrows  indicate the electric field direction. (b)  Snapshots of bead position in a time sequence of proton irradiation at 1.9 MGy s$^{-1}$. The proton beam is centered in the 12 $\mu$m diameter red circle.}
\end{figure}

{\it Experimental set-up.--} 
The experiments are performed using a beamline of the AIFIRA facility of the Laboratoire de Physique des 2 Infinis de Bordeaux~\cite{bourret2014fluorescence}. A 3 MeV proton beam is focused onto a water target at normal incidence as illustrated in Fig. 1(a). The beam intensity varies from 0.25 pA to 12 pA and the fluence distribution has a Gaussian shape with a full width at half maximum of about 5 µm at the target position. This target is a water film confined between 11 µm thick polypropylene sheets. The 3 MeV protons have about 150 µm range in water  losing most of their energy just before coming to rest (the so-called Bragg peak) \cite{pstar}. We use the microscope to check that the target thickness is below 90 – 120 µm to maintain a homogeneous loss of energy ranging from 12 to 17 keV/$\mu$m in water along the proton path in the sample \cite{pstar}. This leads to a cylindrical symmetry of the system allowing a two-dimensional description of the processes at play. Dose rates between $3\times10^4$ to $2\times10^6$ Gy s$^{-1}$ are reached in the target. These values are typical of the dose rates that can be achieved with intense electron or photon micro-beam facilities~\cite{wu2020liquid,ialyshev2022enhancing,eling2019ultra, hemonnot2017imaging}.

The water film contains diluted dye-doped colloidal polystyrene beads with a diameter of 1 µm. These particles are negatively charged by coating their surface with sulfate and carboxylate groups generating a $\zeta$-potential~\cite{lyklema1995solid} measured at $\zeta = -60$ mV in water at pH 6, which is the acidity level before irradiation. To explore the effect of the proton beam on the bead motion, we used an inverted fluorescence microscope. A light beam emitted by a diode with a wavelength of 470 nm is steered through a dichroic mirror into the microscope. An objective illuminates the sample and images the 490 nm bead fluorescence onto a camera. Fig. 1(b) shows a typical example of bead images. The proton beam is switched on at time t=0. The images clearly show beads moving away from the proton beam. Additional information on the beam and target characterizations as well as on the data acquisition and processing are detailed in Sec. I of the Supplemental Material~\cite{supmat}\nocite{crocker1996methods,bernal2015track,incerti2010geant4,ramos2020independent,zeebe2011molecular,olalde2009effect,stumm2012aquatic,pocker1977stopped,elliot2009reaction,hiemenz1986principles}.

{\it Drift velocity.--} 
We measure the radial velocity $u$ of the beads by single-particle tracking. Fig. 2(a) shows velocity profile  in its steady state for proton dose rates ranging from 30 kGy s$^{-1}$ to 1.9 MGy s$^{-1}$. The beads are sensitive to phoretic mechanisms to distance up to r = 100 $\mu$m from the beam axis and the velocities reach 1 to 2 $\mu$m s$^{-1}$ near the proton beam. These profiles are weakly dependent on the dose rate, although it varies by almost two orders of magnitude. The velocity profile tends to decrease slightly more faster than an $r^{-1}$ scaling law. In Fig. 2(b) we report the time evolution of the radial velocity of the beads measured at 3 different distances from the center of the proton beam at 1.9 MGy s$^{-1}$. A transient behavior clearly appears with a duration that increases with distance from the beam axis. 
The transient dynamics is well fit by an exponential law  $u(t)\propto 1-e^{-t/\tau}$, as shown in Fig. 2(b). The time constant varies with the radial distance $r$ according to $\tau=r^2/4D$ (Fig. 2(c)) indicating that a diffusive process is responsible for the bead motion.
The diffusivity $D$ is equal to 2200 $\mu\mathrm{m}^2\mathrm{s}^{-1}$, of the same order of magnitude as the diffusion coefficient of molecules composed of a few atoms in water~\cite{hill1994calculation}. Therefore, these data demonstrate without ambiguity that diffusiophoresis effects near the proton beam cause the particle migration. It excludes the thermophoresis  process~\cite{piazza2008thermophoresis} (thermal diffusivity $\sim$ 1.4 10$^6$ $\mu$m$^2$ s$^{-1}$) as well as non-diffusive mechanisms such as pressure wave effects in the interaction of radiation with water~\cite{toulemonde2009temperature}.

\begin{figure}[t]
\includegraphics[width=8.6cm,trim=0 30 50 210, clip=true]{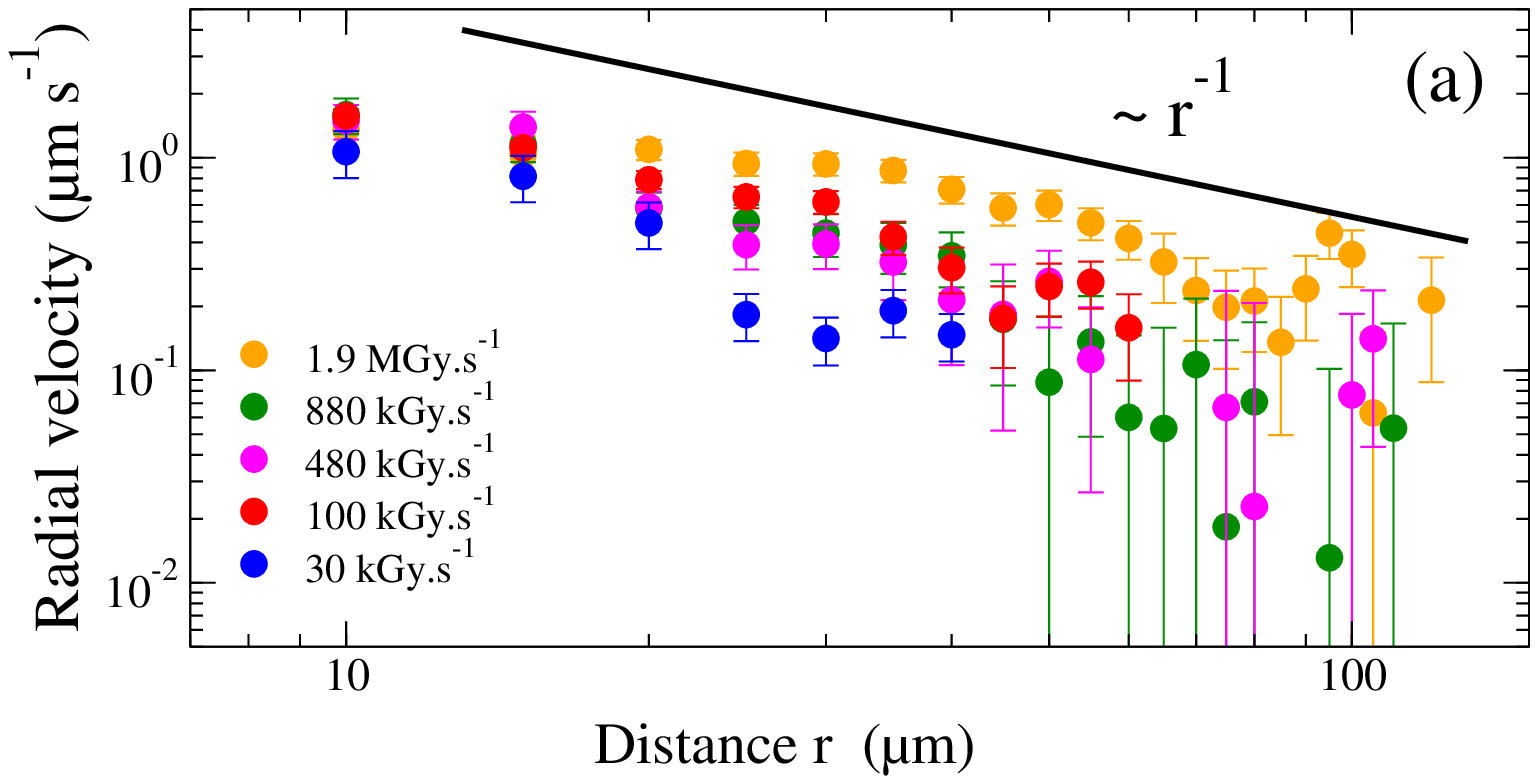}
\begin{minipage}{0.49\linewidth}
\includegraphics[width=4.3cm,trim=20 40 300 230, clip=true]{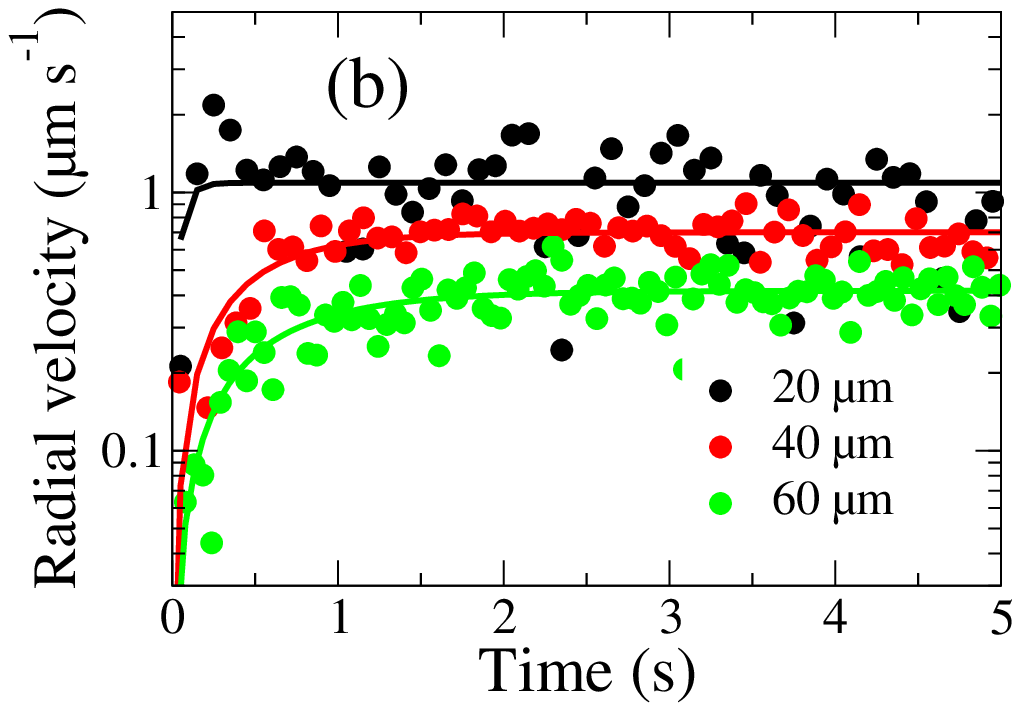}
\end{minipage}
\begin{minipage}{0.49\linewidth}
\includegraphics[width=4.3cm,trim=10 40 280 240, clip=true]{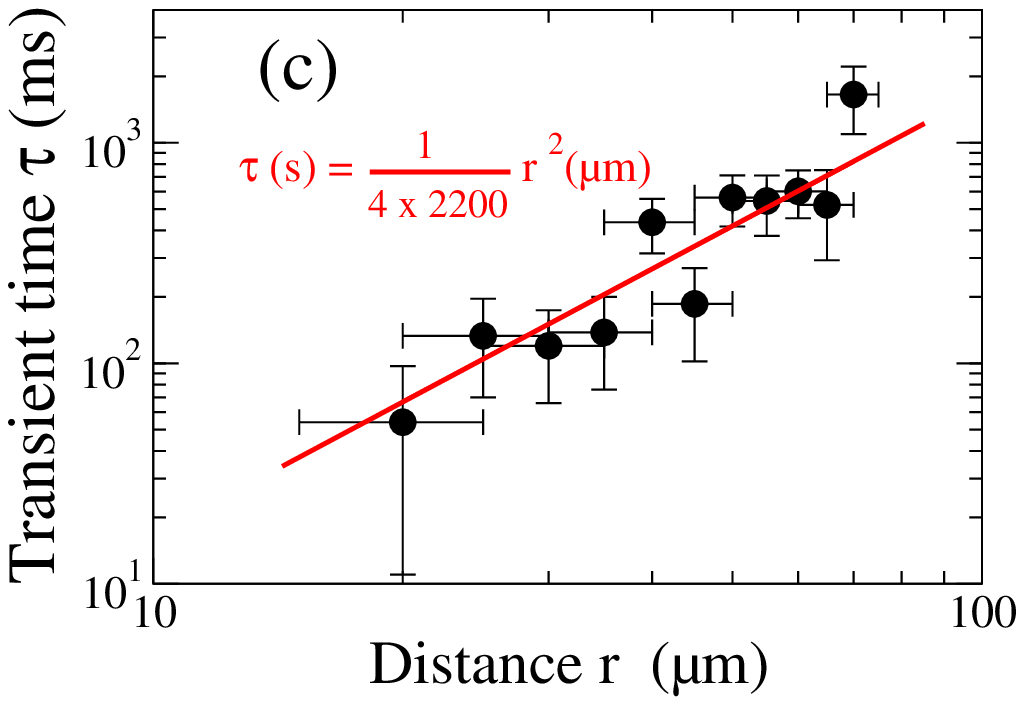}
\end{minipage}

\caption{\label{Fig2} (a) Steady-state radial velocity profiles  measured for proton dose rates ranging from 30 kGy s$^{-1}$ to 1.9 MGy s$^{-1}$. (b) Time evolution of the radial velocity of the beads measured at 3 different distances from the center of the proton beam at 1.9 MGy s$^{-1}$. (c) Evolution of the characteristic time $\tau$ of the transient state as a function of the distance from the beam axis at 1.9 MGy s$^{-1}$.}
\end{figure}

{\it Electric field.--} Molecules created by water radiolysis are either neutral (H$_2$, O$_2$, H$_2$O$_2$, …)  or charged (H$_3$O$^+$,O$_2^-$, OH$^-$, ...) \cite{swallow1973radiation,schneider2014electron}. As neutral solute induced diffusiophoresis is known to be generally much weaker than ion induced one~\cite{derjaguin1947kinetic,anderson1982motion,anderson1989colloid,anderson1984diffusiophoresis,shim2022diffusiophoresis,ajdari2006giant,abecassis2008boosting,paustian2015direct}, in the following, we assume that the migration is related to 
ion gradients in the solution.  
This scenario will be supported later by numerical calculations. 
The hydronium ion H$_3$O$^+$ has the highest diffusivity $D_+$ = 9000 $\mu$m$^2$ s$^{-1}$ whereas anion diffusivities range from 1000 to 3000 $\mu$m$^2$ s$^{-1}$. This diffusivity contrast results in charge separation, with an excess of anions in the beam region and a corresponding excess of cations at the outer boundary, while the bulk solution remains neutral. Accordingly, there is an inward electric field as shown in Fig. 1a.

To estimate the magnitude of the electric field, we first describe this system as a simple model equivalent to a 1:1 electrolyte consisting of cations H$_3$O$^+$ and average anions A$^-$ with the concentration field in the steady state $c(r)=c_{H_3O^+}(r)=c_{A^-}(r)$. As we will see later, the relevant anions are O$_2^-$ and HCO$_3^-$ with respective diffusivities of 2100 and 960 $\mu$m$^2$ s$^{-1}$ so that $D_{A^-} = D_-$ = 1500 $\mu$m$^2$ s$^{-1}$ is a reasonable value.  The fact that no current arises in the solution, together with its electroneutrality, implies the creation of an electric field ~\cite{prieve1984motion}: 

\begin{equation}
{\Eb}_{1:1}=\frac{k_BT}{e}\frac{D_+-D_-}{D_++D_-} {\nabla} \ln (c) 
\end{equation}
with $k_BT$ the thermal energy and $e$ the elementary charge. 
The migration velocity of a particle with $|\zeta| <$ 75 mV  in the ion concentration gradient is~\cite{prieve1984motion}:

\begin{equation}
\ub =\frac{\epsilon\zeta}{\eta}\left(\Eb_{1:1}   +\frac{\zeta}{8} \nabla \ln(c) \right),
\end{equation}

where the first term is the electrophoretic contribution depending on the $\zeta$-potential, the permittivity $\epsilon$, and the viscosity $\eta$. The second term is the weaker chemiophoretic component induced by the interactions of ions with the charged particles which push the beads towards higher ion concentrations~\cite{anderson1989colloid,anderson1984diffusiophoresis,shim2022diffusiophoresis}.

Given the particle migration velocities involved, it is interesting to estimate an order of magnitude of the expected concentrations of cations and anions. Typical parameter values are $k_BT/e \approx$ 27 mV, and $u$ $\approx$ 1 $\mu$m s$^{-1}$ thus $\nabla \ln(c)$ is of the order of 20 cm$^{-1}$. This value corresponds to a 10\% decrease in the H$_3$O$^+$ concentration in about 40 $\mu$m. Considering an initial solution at pH 6, we conclude that the H$_3$O$^+$ concentration near the beam is at most a few $\mu$M and that the pH changes locally by only a few tenths of a unit. 
We can then consider that the beads have the same $\zeta$-potential at any point in the solution. According to equations (1) and (2), the amplitude of the electric field in the steady state can be extracted from the experimental values of $u_{\text{exp}}(r)$ at a given distance $r$ as:
\begin{equation}
\Eb_{\text{exp}}= 
  \frac{\eta}{\epsilon \zeta}\frac{\beta} {\beta + (e\zeta/8k_BT)} \ub_{\text{exp}}  
\end{equation}

with $\beta = \frac{D_+-D_-}{D_++D_-}$.
The estimated experimental values of the electric field are calculated from this equation and are reported in Fig. 3 considering $\zeta$ = $-60$ mV. The field strengths decrease monotonically with the distance from the beam axis between 10 and 120 $\mu$m and have a maximum value between 50 and 70 V m$^{-1}$ depending on the dose rate. Electric fields of the order of 10 V m$^{-1}$ extend to distances greater than 100 $\mu$m from the beam. They are much stronger than the electric fields that are directly generated by the charges of the beam protons themselves.

\begin{figure}

\includegraphics[width=7cm,trim=10 40 220 230, clip=true]{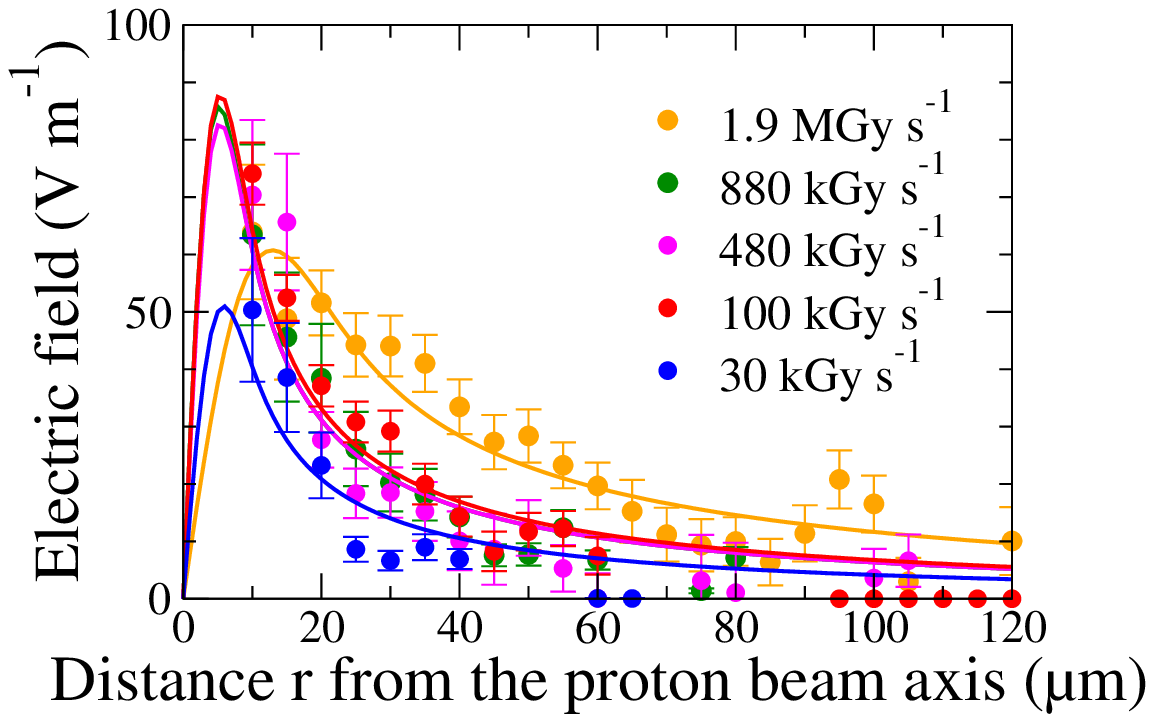}

\caption{\label{Fig3} Profiles of the magnitude of the electric field extracted from the particle migration velocities measured in their steady-states for various proton dose rates. Symbols are experimental data and solid lines are fits according to equations (1) and (4). 
}
\end{figure}


We now determine if the profiles of electric fields are compatible with a simple scenario in which cations and anions are produced within the beam radius, at a rate $K_0 S(r)$, where $S(r)=e^{-r^2/(2\sigma^2)}/(2\pi\sigma^2)$ is a normalized Gaussian of variance $\sigma^{2}$. Although both cations and anions are subject to the electric field that they generate, in the electroneutrality approximation their distribution obeys a (force free) diffusive dynamics with an effective diffusivity $D = \frac{2D_+ D_-}{D_+ + D_-}$ = 2600 $\mu$m$^2$ s$^{-1}$ (see Sec. II.A of the Supplemental Material \cite{supmat}). It is then clear that the profile at a time $t$ is:

\begin{equation}
c(r,t)=c(r,0) + \int_0^t K_0 e^{-\frac{r^2}{2\sigma^2+4D\tau}}\frac{1}{2\pi (\sigma^2+2D\tau)}d\tau
\end{equation}

from which we can obtain the expression for the electric field via Eq. (1). We consider $c(r,0)$ = 1 $\mu$M, corresponding to pH = 6, whereas the two free parameters K$_o$ and $\sigma$ are chosen to minimize the residual $\sum (E_{1:1}(r)-E_{exp}(r))^2$. 
The calculations are carried out for the irradiation time t = 5 s representative of a steady state and are plotted with a solid line in Fig. 3. We obtain good agreement between the analytical law and the experimental measurements with $\sigma$ $\sim$ 4 to 8 $\mu$m and  $K_0$ $\sim$ 250 to 900 $\mu$M $\mu$m$^{2}$ s$^{-1}$ depending on the dose rate. 
The predicted profiles are thus compatible with experimental ones, at the cost of introducing an effective emission rate $K_0$ with a  strongly non-linear dependence on the power-beam, possibly representing the effect of destruction of species by chemical reactions.

{\it  Chemical reaction model.--}
In the second part of this letter we support the scenario of ion-induced diffusiophoresis 
by using a reaction-diffusion model introduced in Ref.~\cite{schneider2014electron}. We modify this model to take into account the migration of ions in the electric field they generate and we consider a few additional chemical species. More precisely, we assume that the evolution of the concentration c$_i$ of species $i$ is given by the \textcolor{blue} {
transport} equation: 

\begin{eqnarray}
\frac{\partial c_i}{\partial t} = -\nabla\cdot\Jb_i-R^i_d+R^i_p+S^i_g
\end{eqnarray}
with $\Jb_i$ the flux of species $i$, $R^i_d$ and $R^i_p$ their rate of destruction and production by aqueous chemical reactions, and $S^i_g$ its rate of generation by incident protons. The flux of chemical species is given by the Nernst-Planck equation $\Jb_i = D_i(-{\nabla} c_i + \frac{z_ie}{k_BT} c_i \Eb) $ with $D_i$ the diffusivity of chemical species $i$.
The first term of the flux is Fick's law whereas the second term describes the drift component of the ionic species with electric charges z$_i$e due to the electric field $\Eb$ produced in the solution. The amplitude of this field is calculated at a given time and position using the integral form of Gauss' law with the concentration profiles of the charged species. The generation rate $S^i_g$ is related to the yields of the chemical species $i$ produced after the energy deposition of 100 eV in water by an incident proton. These yields, named $g_i^o$-values, are calculated with the Monte Carlo toolkit Geant4-DNA~\cite{incerti2018geant4}. 
The rates of chemical reactions are taken from Ref.~\cite{schneider2014electron}, except for the additional chemical species we add, see Sec. II.B of  the Supplemental Material~\cite{supmat} for details.

Figure 4(a) shows the spatial behavior of some of the expected chemical species in the solution, 5 s after the proton beam is switched on at a dose rate of 1.9 MGy s$^{-1}$. The most reactive radiolysis products (i.e. for which $R^i_d \sim R^i_p + S^i_g$), such as H$_3$O$^+$ or O$_2^-$, diffuse from the irradiated region, where they are produced, and are rapidly consumed by chemical reactions.  Weaker reacting species (i.e. for which $R^i_d$ and $R^i_p \ll  S^i_g$), such as H$_2$O$_2$ or H$_2$, can diffuse further in the target slowly increasing their concentration throughout the liquid cell. The concentration gradients are all directed towards the proton beam, except for HCO$_3^-$, which is the only chemical species with a  minimum concentration near the beam. This molecule is not produced during radiolysis but is consumed by H$_3$O$^+$ ions to reach a chemical equilibrium with H$_2$CO$_3$ and CO$_2$. Finally, only three ionic species contribute mainly to the production of the electric field: H$_3$O$^+$, O$_2^-$ and HCO$_3^-$. This configuration is close to the 1:1 electrolyte model considered to extract the electric field in Fig. 3.

\begin{figure}

\includegraphics[width=7.5cm,trim=0 10 250 20, clip=true]{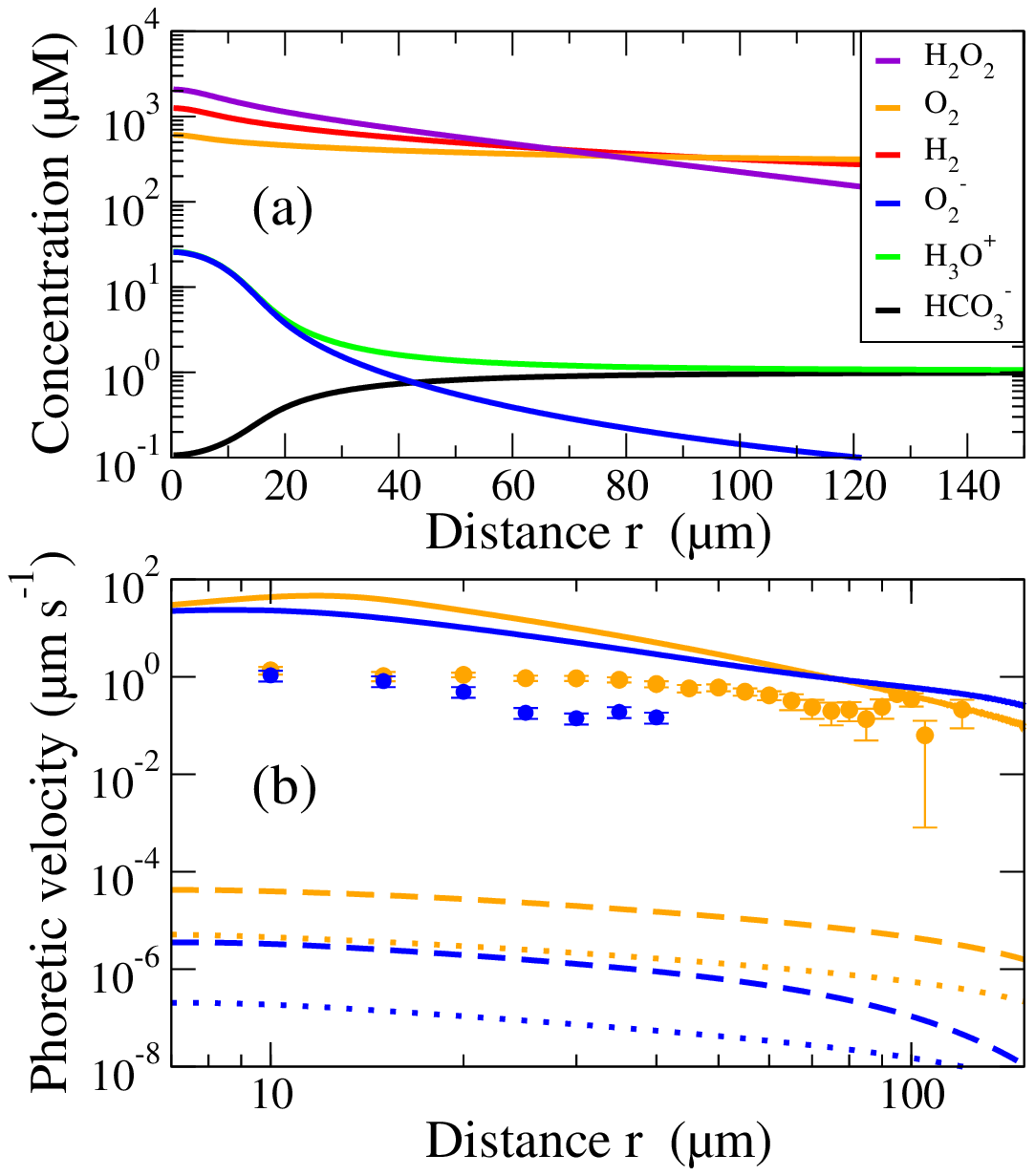}

\caption{\label{Fig4} (a) Concentration profiles at t = 5 s. Calculations are performed with the reaction-diffusion model at a dose rate of 1.9 MGy s$^{-1}$. (b) Experimental (symbol) and calculated bead migration velocities at dose rates of 30 kGy s$^{-1}$ (blue) and 1.9 MGy s$^{-1}$ (orange). Calculated velocities induced by polar neutral (eq. 6), apolar neutral (eq. 7) and charged (eq. 8) species gradients are reported in dashed, dotted and solid lines, respectively.}
\end{figure}

{\it Drift velocity predictions.-- } 
We now estimate the migration velocities of the beads in this reaction-diffusion model. Fig. 4(a) shows that the strongest concentration gradients are obtained for the neutral species H$_2$O$_2$, H$_2$ and O$_2$ 
which induce diffusiophoresis driven by the weak ion-dipole interactions 
 between the charged particles and neutral solutes ~\cite{derjaguin1947kinetic}. In the case of polar neutral (pn) molecules H$_2$O$_2$ with dipole moment $\mu_D$  $\sim$ 1.6 D, the migration velocity of the particles induced by the concentration gradient $\nabla$c$_{\text{pn}}$, is given by~\cite{anderson1989colloid}:
\begin{equation}
\ub_{\text{pn}}=\frac{-1}{12\eta}\frac{\mu_D^2}{k_BT}\zeta^2 \nabla c_{\text{pn}},
\end{equation}
whereas for apolar neutral (an) molecules such as H$_2$ or O$_2$ of volume   $V\sim$ 10$^{-30}$ m$^3$ the 
velocity 
is related to the concentration gradient $\nabla$c$_{\text{an}}$ as:
\begin{equation}
\ub_{\text{an}}=\frac{-3V\epsilon}{12\eta}\zeta^2 \nabla c_{\text{an}}.
\end{equation}
The negative sign means that $u_{pn}$ and $u_{an}$ are directed towards lower solute concentrations.
The calculated velocity profiles are reported for these two contributions in Fig. 4(b) for dose rates 30 kGy s$^{-1}$ and 1.9 MGy s$^{-1}$.  
With typical parameter values, they are of the order of $10^{-7} - 10^{-5}$ µm s$^{-1}$ which is negligible compared to the observed migration velocities, indicating that 
the origin of the bead migration cannot be attributed to neutral solute gradients.

Finally, with gradients of more than two ionic species, the expression of the bead migration velocity is still composed of the electrophoretic and chemiophoretic components as: ~\cite{chiang2014multi}:
\begin{eqnarray}
\ub=\zeta\frac{\epsilon k_BT}{e\eta}\frac{\sum z_i D_i \nabla c_i}{\sum z_i^2 D_i c_i}
+ \zeta^2\frac{ \epsilon }{8\eta}\frac{\sum z_i^2 \nabla c_i}{\sum z_i^2 c_i},
\end{eqnarray}

that gives the expression (2) in the case of a 1:1 electrolyte. 
Fig. 4(b) shows the calculated profiles. The expected velocities are five to seven orders of magnitude larger than the previous ones, which unambiguously justifies ion-induced diffusiophoresis as the main effect observed in the experiment.
Note that without any fitting parameter, this model clearly predicts velocity profiles  similar to the experiment, and in particular the weak dependency of the velocities on the dose rate. The discrepancy of about one order of magnitude between simulations and experiments could be explained by various reasons. In the case of H$_3$O$^+$ ions, the strong competition between the rate of generation $S^i_g$ and the rate of destruction $R^i_d$ by chemical reactions makes a quantitative estimation of its concentration profile very difficult. Indeed, for these species, the production and destruction rates differ only by less than $0.01 - 0.1 \%$, so that an accurate prediction for the concentration of these ions is difficult (see Supplementary Material, Section 2C). Some chemical reactions have kinetic orders different from one, leading to non-linear effects. This is probably the reason for the weak dependence of the particle velocity on the dose rate. Moreover, the g$^0$-values and the reaction rates extracted from low dose rate studies probably cannot be simply considered for physico-chemical processes at dose rates above 10 kGy s$^{-1}$. Indeed, in our experiments, these rates are several orders of magnitude higher than those considered by the reactor physics community, from which the code is derived~\cite{elliot1990computer}. 
Nevertheless, although our calculations overestimate the migration velocities, they make clear that the generation of an electric field as a consequence of the production of ions by water radiolysis is plausible.

In summary, we have experimentally observed the migration of colloidal beads in a solution irradiated by a proton beam. Our observations of transient regimes point towards electrophoresis as responsible for bead migration. A theory taking into account the creation and reactions of molecules  rules out the possibility that migration is due to neutral molecules, leading to the conclusion that the migration of the beads is linked to an electric field generated by a non-uniform density of charged molecules. The migration velocities predicted by our theory, although overestimated, share many qualitative features with our experimental observations.

The study has been performed with proton beams but the conclusions are relevant to focused beams of other ionizing radiations. With 300 keV electron beams used in transmission electron microscopy (TEM) in the liquid phase~\cite{wu2020liquid,schneider2014electron} and dose rates of a few tens of MGy s$^{-1}$, similar electric fields are expected at 20 $\mu$m from the axis of the electron beam. This raises the question of the disturbance near the electron beam in biological or soft matter systems studied by TEM in liquid phase. 

In this study, the proton beam does not move and the electric field generated by the ion concentration gradients is radial and static after a transient state. The proton beam can be easily manipulated with the electrostatic elements of the experimental set-up. The beam can thus be alternately applied to different regions of the target over short or long duration relative to the characteristic diffusion times of the solutes in water. In the presence of several moving sources of ion gradients, much more complex electric field lines can be generated to trap or manipulate, for example, one or a few colloidal particles or macromolecules. There is also no difficulty in irradiating the target with an array of regular or random sources opening up possibilities to study particle dynamics in complex two-dimensional random landscapes~\cite{evers2013particle}.

\begin{acknowledgments}

This project has received financial support from the CNRS through the MITI interdisciplinary programs.

The AIFIRA facility is financially supported by the CNRS, the University of Bordeaux and the Région Nouvelle Aquitaine. We thank the technical staff
members of the AIFIRA facility P. Alfaurt, J. Jouve and S. Sorieul.

We thank Dr. M. Toulemonde for helpful discussions and Dr. D. Smith for careful reading of the manuscript.
\end{acknowledgments}

\bibliography{apssamp}

\end{document}